\documentclass{revtex4}
\usepackage{amsmath}
\usepackage{amsfonts}
\usepackage{amssymb}
\usepackage{graphicx}

\begin{document}

\title{Measurements on Composite Qudits}

\author{Tomasz Paterek}
\email{pater@univ.gda.pl}
\affiliation{Institute of Theoretical Physics and Astrophysics, University of Gda\'nsk, Poland}
\affiliation{The Erwin Schr\"odinger International Institute for Mathematical Physics,
Boltzmanngasse 9, A-1090 Vienna, Austria}

\begin{abstract}
We study measurements of the unitary generalization of Pauli operators.
First, an analytical (constructive) solution to the eigenproblem of these operators is presented.
Next, in the case of two subsystems,
the Schmidt form of the eigenvectors
is derived to identify measurements which are easy to implement.
These results are utilized to show that quantum cryptography
with two bases, when operating on a two-component qudit, 
can be realized with measurements on individual subsystems,
assisted with classical communication.
We also discuss feasible devices 
which perform tomography of 
polarisation-path qudits.
\end{abstract}


\maketitle

\section{Introduction}

Compared to qubits,
higher-dimensional quantum systems 
improve performance of many protocols and algorithms
of quantum information processing.
For example, additionally to their increased capacity,
they make quantum cryptography more secure \cite{1PERES,1BRUS,MOHAMED},
or lead to a greater reduction of communication complexity \cite{PRL_CCP,IJQI_CCP}.
One way to deal with a qudit is to find a convenient physical system representing it.
A beautiful example is a photon with many accessible propagation paths \cite{1MULTIPATH,2MULTIPATH}.
Another approach, studied here, is to treat many systems of lower dimensions
as a global higher-dimensional object -- a composite qudit.
The challenge is to prepare and operate on
entangled states of subsystems
and to experimentally realize all global observables.
This usually requires difficult conditional operations.

The preparation of entangled states (at least some of them)
is well within a reach of current technology.
For example, entanglement of two photons,
in all degrees of freedom, was demonstrated in Ref. \cite{HYPERENTANGLEMENT}.
Each of these photons can be regarded as a composite qudit,
with subsystems represented by different degrees of freedom.
Further, a system of two photons can be thought of as an even higher
dimensional qudit.

Here, measurements on such qudits are studied.
Some examples are already proposed
and realized in a context of Bell's theorem \cite{PROPOSAL,1EXP,2EXP}.
The present paper is a generalization of that work.
First, the requirements for an arbitrary global observable
are given.
Next, a specific class of operators, unitary generalizations
of Pauli operators \cite{SCHWINGER,FIVEL,GOTTESMAN,PR}, is described in detail.
The importance of this class comes from its applications.
For example, the operators form a full tomographic set 
(allow for a reconstruction of a density matrix),
appear in quantum cryptography \cite{1PERES,1BRUS,MOHAMED},
or tests of local realism \cite{PROPOSAL,1EXP,2EXP,1BELL,2BELL,3BELL}.
A solution to the eigenproblem of these operators is constructed in full generality
and, for a two-component case,
the Schmidt representation of the eigenstates is derived.
All the eigenvectors are shown to have the same Schmidt number.
Thus, the sets of entangled and disentangled eigenbases are identified,
which respectively define the sets of ``more difficult'' and ``easier'' 
realizable operators.
A beautiful method to solve the eigenproblem
of the generalized Pauli operators, 
based on Euclid's algorithm,
was given by Nielsen \emph{et al.} \cite{NIELSEN}.
However, since their interests were different,
they did not present an explicit solution.
The generalized Pauli operators
have been studied in various levels of detail
in many other papers.
Nevertheless, the present author
could not find a general form of the eigenbasis.
Here, an explicit compact formulae
for eigenvectors and eigenvalues are given,
as well as a practical procedure how to compute them.

Although unitary, the generalized Pauli operators are measurable.
In quantum mechanics, 
different outcomes of a measurement apparatus correspond to
different orthogonal states of a system.
Due to the fact that most often measurement outcomes
are expressed in form of real numbers we are used to connect 
Hermitian operators with observables.
However, there are measurement apparatuses which \emph{do not} output a number.
Take a device which clicks if a photon is detected
or a bunch of such photo-detectors which monitor
many possible propagation paths of a photon.
The operator
associated with this apparatus has a specific spectral decomposition
(different clicks find the system in different orthogonal states).
However, the eigenvalues assigned to the clicks can be arbitrary,
as long as the assignment is consistent,
i.e. clicks of the same detector always reveal the same eigenvalue.
If one finds it useful to work with complex eigenvalues,
as it is often the case when considering higher-dimensional quantum systems,
one can use operators which are unitary,
with eigenvalues given by the complex roots of unity.

With any generalized Pauli operator 
one can associate a measurement device capable to measure it.
We present such devices for polarisation-path qudits,
and prove that quantum cryptography 
with two bases 
is relatively easy to realize as it
does not require any joint measurements on the subsystems.
As the unitary operators correspond to certain measurement apparatuses,
they will be often called ``observables''.

\section{General measurements}

Consider a qudit composed of many subsystems, possibly of different dimensions.
The measurement of any global observable can be viewed as a unitary
evolution of the whole system which transforms the eigenvectors of the observable 
into the eigenvectors which can be distinguished by the measurement apparatus.
For subsystems of equal dimensions arbitrary global unitary operation can be decomposed into
local and two-body conditional operations \cite{SPECTRAL_QUDITS}.
This proof can be almost directly applied to the problem studied here,
and it will not be repeated.
Individual measurements, local and conditional two-body operations
are sufficient to realize any global measurement on a composite qudit.

Instead of finding the evolution,
one can decompose a global observable
into (possibly joint) measurements on subsystems
and classical communication.
Eigenbases of individual measurements
form a product basis in a global Hilbert space.
Eigenvectors of any global observable
can be decomposed in this basis.
If the eigenvectors 
factorize, that is 
$| j \rangle = | j_{N-1} \rangle_{N-1} |j_{N-2} \rangle_{N-2} ... |j_{0} \rangle_0$,
where $|j_{n} \rangle_n$ is a state of subsystem $n$,
then there are two possible scenarios:
(A) a global measurement can be performed with
individual measurements on separate subsystems,
(B) it can be done with an additional use of a feed-forward technique,
i.e. a subsequent measurement setting depends on the outcomes of all previous measurements.
To see this, note that
orthogonality of vectors $| j \rangle$ implies certain orthogonalities
of the states of subsystems, $| j_n \rangle_n$.
In the simplest case, for each subsystem the vectors $| j_n \rangle_n$ form a basis.
Then, the first scenario, (A), can be applied.
The other possibility is that vectors, say $| j_0 \rangle_0$,
form an orthogonal basis, and for every $| j_0 \rangle_0$
one has a \emph{different} set of orthogonal vectors of another subsystem, 
say $| j_1 \rangle_1$, and so on.
In this case, one first measures the particle
the states of which span the full basis (in our case subsystem ``0'').
Next, depending on the outcome,
another subsystem is measured in a suitable basis.
Further on, depending on both previous outcomes, yet another subsystem is measured, etc.
This is what is called feed-forward technique, (B).
If some eigenstates of a multisystem observable
do not factorize,
joint measurements are necessary to measure it.

\section{Eigenproblem of the generalized Pauli operators}

In any case, the realisation of a global observable is
based on the solution of its eigenproblem.
Here, a general solution to the eigenproblem
of the generalized Pauli operators
is presented.

In the Hilbert-Schmidt space of operators
acting on vectors in a Hilbert space of dimension $d$,
one can always find a basis set of $d^2$ unitary operators.
It has been shown that one can construct such a set
using the following relation \cite{SCHWINGER,FIVEL,GOTTESMAN,PR}:
\begin{equation}
S_{kl} = S_x^k S_z^l \quad \textrm{with} \quad k,l=0,...,d-1,
\end{equation}
where the action of the two operators on the right-hand side,
on the eigenvectors of $S_z$ operator, $| \kappa \rangle_z$, is defined by:
\begin{eqnarray}
S_z |\kappa \rangle_z &=& \alpha_d^{\kappa} |\kappa \rangle_z, \label{S_DEFINITION} \nonumber \\
S_x |\kappa \rangle_z &=& |\kappa + 1 \rangle_z, \quad \textrm{where} \quad \kappa=0,1,...,d-1,
\end{eqnarray}
with
\begin{equation}
\alpha_d = e^{i 2\pi/d}.
\label{ALPHA}
\end{equation}
The number $\alpha_d$ is a primitive complex $d$th root of unity,
whereas the addition, here $\kappa+1$, is taken modulo $d$.
Unless explicitly stated all additions are taken modulo $d$.
The operators $S_{kl}$ are called generalized Pauli operators
as for $d=2$ they reduce to standard Pauli operators.
They share some features with them \cite{PR}.
The matrix of any $S_{kl}$, written in the $S_z$ basis,
has only $d$ non-vanishing entries, one per column and row:
\begin{equation}
S_{kl} = 
\left(
\begin{array}{ccccccc}
0      & 0        &        & 0                 & \alpha_d^{(d-k)l} &        & 0 \\
\vdots & \vdots   &        & \vdots            & \vdots          &        & \vdots \\
0      & 0        &        & 0                 & 0               &        & \alpha_d^{(d-1)l} \\
1      & 0        & \ldots & 0                 & 0               & \ldots & 0\\
0      & \alpha_d^l &        & 0                 & 0               &        & 0\\
\vdots & \vdots   &        & \vdots            & \vdots          &        & \vdots \\
0      & 0        &        & \alpha_d^{(d-k-1)l} & 0               &        & 0
\end{array}
\right).
\end{equation}
The only non-vanishing element of the first column, a ``1'',
appears in the $k$th row (recall that $k=0,1,...,d-1$).
Generally, the matrix elements of $S_{kl}$ operator,
$[S_{kl}]_{rm}$, are given by 
$[S_{kl}]_{rm} = \delta_{r-k,m} \alpha_d^{ml}$,
where $\delta_{x,y}$ is the Kronecker delta.
Since every $S_{kl}$ is unitary it can be diagonalized:
\begin{equation}
S_{kl} = V D V^{\dagger},
\label{DIAGONALIZATION}
\end{equation}
where $V$ 
is a unitary matrix the columns of which are eigenstates of $S_{kl}$, 
$V = (|0 \rangle, ..., |d-1 \rangle)$,
and $D$ is a diagonal matrix with entries being eigenvalues of $S_{kl}$, denoted by $\lambda_{j}$.
The form of $[S_{kl}]_{rm}$ and  (\ref{DIAGONALIZATION}) imply 
conditions, which must be satisfied by the eigenvectors $|j \rangle$:
\begin{equation}
\sum_{j = 0}^{d-1} \lambda_{j} v_{k+m,j} v_{m,j}^{*} = \alpha_d^{ml}, \quad \textrm{for all} \quad m=0,...,d-1,
\label{CONDITION}
\end{equation}
where $v_{i,j}$ is the element of the matrix $V$ in the $i$th row and $j$th column, 
i.e. the $i$th coefficient of the eigenvector $|j \rangle$.
A study of this condition allows one to construct the eigenbasis.

We first present the result, that is give a candidate for an eigenbasis,
and then prove that this is indeed the eigenbasis.
Depending on $k$ ,
the eigenstates of $S_{kl}$ are given 
by superposition of different number of states $| \kappa \rangle_z$.
Let us denote by $f$ the 
greatest common divisor of $k$ and $d$.
Within this definition $k=wf$ is a multiple of $f$.
The eigenstates $| j \rangle$
involve every $f$th state of the $S_z$ basis:
\begin{equation}
| \kappa \rangle_z = |a + \eta' f \rangle_z = |a + \eta k \rangle_z,
\label{STATES_INVOLVED}
\end{equation}
where $\eta'=0,...,d/f-1$ and $a=0,...,f-1$,
and of course $\eta' = w \eta$.
Both $\eta'$ and $\eta$ enumerate different states $| \kappa \rangle_z$
into which $| j \rangle$ is decomposed, i.e. $\eta=0,...,d/f-1$.
All other coefficients vanish.
The whole eigenbasis splits into $f$
groups of eigenvectors which are superpositions of vectors $|\kappa \rangle_z$
with fixed $a$.
There are $d/f$ eigenvectors within each group.
To uniquely identify the eigenvector $| j \rangle$
one needs to specify $a$, and additionally an integer $g=0,...,d/f-1$, i.e. $j = j_{g,a}$.
With these definitions we can present
the form of eigenvectors (a candidate):
\begin{equation}
|j \rangle = |j_{g,a} \rangle = \frac{1}{\sqrt{d/f}}
\sum_{\eta=0}^{d/f-1} \lambda_{j_{g,0}}^{- \eta} \alpha_d^{\frac{ \eta (\eta -1)}{2}kl} | a + \eta k \rangle_z,
\label{EIGENVECTORS}
\end{equation}
where generally the eigenvalues $\lambda_{j_{g,a}}$ are given by:
\begin{equation}
\lambda_{j_{g,a}} = e^{i \varphi} \alpha_d^{gf+al},
\label{EIGENVALUES}
\end{equation}
and $e^{i \varphi}$ is a phase factor common to all the eigenvalues.\footnote{To get rid of this phase, instead of $S_{kl}$
one can consider an operator $e^{-i \varphi} S_{kl}$.}
We will show below how to compute this phase.
Note that the coefficients in (\ref{EIGENVECTORS}) are independent of $a$.
This can be intuitively explained by noting
that for different $a$'s the eigenvectors $|j_{g,a} \rangle$
are orthogonal just due to the fact that they involve 
orthogonal vectors $| a + \eta k\rangle_z$.
For a fixed $a$, but different $g$'s, the vectors (\ref{EIGENVECTORS}) 
are also orthogonal.
Their scalar product
$\langle j_{g',a} | j_{g,a} \rangle = (d/f)^{-1} \sum_{\eta = 0}^{d/f-1}(\lambda_{j_{g',0}} \lambda_{j_{g,0}}^{-1})^{\eta}$
involves the product of 
$\lambda_{j_{g',0}} \lambda_{j_{g,0}}^{-1} = \alpha_d^{(g'-g)f} = \alpha_{d/f}^{(g'-g)}$,
and the whole sum is equal to Kronecker delta $\delta_{g',g}$.
Thus, the vectors $| j_{g,a} \rangle$ form an orthonormal basis.

To prove that this basis is the eigenbasis one needs to check whether
\begin{equation}
S_{kl} |j_{g,a} \rangle = \lambda_{j_{g,a}} |j_{g,a} \rangle.
\label{EIGEN_JA}
\end{equation}
The action of $S_{kl}$, defined by (\ref{S_DEFINITION}), on the state $|j_{g,a} \rangle$ is given by:
\begin{equation}
S_{kl} |j_{g,a} \rangle =
\frac{1}{\sqrt{d/f}} \sum_{\eta =0}^{d/f-1} \lambda_{j_{g,0}}^{- \eta} \alpha_d^{\frac{ \eta (\eta -1)}{2}kl} 
\alpha_d^{l(a+ \eta k)}| a+(\eta+1)k\rangle_z,
\end{equation}
Changing the summation index to $\eta_1 = \eta+1$
one finds:
\begin{equation}
S_{kl} |j_{g,a} \rangle = 
\lambda_{j_{g,0}} \alpha_d^{al} 
\frac{1}{\sqrt{d/f}} \sum_{\eta_1=1}^{d/f} \lambda_{j_{g,0}}^{-\eta_1} \alpha_d^{\frac{\eta_1(\eta_1-1)}{2}kl} 
| a+\eta_1 k\rangle_z.
\label{S_J}
\end{equation}
The coefficients within the sum are equal to the coefficients of the initial
$| j_{g,a} \rangle$ state 
if for the last term in (\ref{S_J}),
for which $\eta_1 = d/f$, one has:
\begin{equation}
\lambda_{j_{g,0}}^{-d/f} = \alpha_d^{-\frac{1}{2}\frac{d}{f}(\frac{d}{f}-1)kl}.
\label{COMPUTE_EIGENVALUES}
\end{equation}
This equation gives the eigenvalues $\lambda_{j_{g,0}}$.
If one takes one of the solutions to (\ref{COMPUTE_EIGENVALUES}), say $\lambda_{j_{0,0}}$,
in the form $\lambda_{j_{0,0}} = e^{i \varphi}$, then the remaining solutions
are given by $\lambda_{j_{g,0}} = e^{i \varphi} \alpha_{d/f}^g$.
Indeed, if $\lambda_{j_{0,0}}$ satisfies (\ref{COMPUTE_EIGENVALUES}),
then also $\lambda_{j_{g,0}}$ satisfy it.
The eigenvalues for other $a$'s are given by:
\begin{equation}
\lambda_{j_{g,a}} = \lambda_{j_{g,0}} \alpha_{d}^{al}.
\label{LAMBDA_A_DEF}
\end{equation}
Note that degeneracies in the eigenproblem can only appear for $f \ne 1$.
(since for $f=1$ one has only $a=0$, and $g$ takes all $d$ different values).

The eigenvalues of $S_{kl}$ operator are
rotated in the complex plane by $e^{i \varphi}$
from the complex roots of unity.
If one puts $e^{i \varphi} = \alpha_d^x$
to be some power of $\alpha_d$,
from Eq. (\ref{COMPUTE_EIGENVALUES})
this power is given by $x = \frac{kl}{2}(\frac{d}{f}-1)$.
Thus, the eigenvalues of operator $\alpha_d^{-\frac{kl}{2}(\frac{d}{f}-1)} S_{kl}$
are rotated back to the complex roots of unity,
which can be a useful property.

Practically, to compute the eigenvectors one should find the value of $f$.
If it is different than unity, set $a=0$ and compute
the coefficients according to Eq. (\ref{EIGENVECTORS}).
For other values of $a$ the coefficients are the same,
but now they multiply orthogonal vectors $| a + \eta k\rangle_z$.
To compute the eigenvalues of $S_{kl}$ one needs to solve Eq. (\ref{COMPUTE_EIGENVALUES}).
It has the following solutions: $\lambda_{j_{0,0}} = \alpha_d^{\frac{kl}{2}(\frac{d}{f}-1)}$;
The other eigenvalues for $a=0$ 
are obtained by multiplication of $\alpha_{d/f}$;
The eigenvalues for $a \ne 0$ can be found from (\ref{LAMBDA_A_DEF}).

\emph{Example}.
Take $S_{43}$ for $d=6$, i.e. $k=4, l=3$ and one finds $f=2$.
Put $a=0$.
From (\ref{COMPUTE_EIGENVALUES}) one has 
$\lambda_{j_{g,0}} = e^{ig\frac{2\pi}{3}} = \alpha_6^{2g}=\alpha_3^{g}$ ($e^{i \varphi}=1$).
According to (\ref{LAMBDA_A_DEF}),
the eigenvalues $\lambda_{j_{g,1}}$ are equal to $\lambda_{j_{g,1}} = -\lambda_{j_{g,0}}$.
This can be summarized in the eigenbasis:
\begin{eqnarray}
|0 \rangle &=& (1/\sqrt{3})\Big[|0\rangle_z + |2\rangle_z + |4\rangle_z \Big], \nonumber \\
|1 \rangle &=& (1/\sqrt{3})\Big[|1\rangle_z + \alpha_3^2 |3\rangle_z + \alpha_3 |5\rangle_z \Big], \nonumber \\
|2 \rangle &=& (1/\sqrt{3})\Big[|0\rangle_z + \alpha_3 |2\rangle_z + \alpha_3^2 |4\rangle_z \Big], \nonumber \\
|3 \rangle &=& (1/\sqrt{3})\Big[|1\rangle_z + |3\rangle_z + |5\rangle_z \Big], \nonumber \\
|4 \rangle &=& (1/\sqrt{3})\Big[|0\rangle_z + \alpha_3^2 |2\rangle_z + \alpha_3 |4\rangle_z \Big], \nonumber \\
|5 \rangle &=& (1/\sqrt{3})\Big[|1\rangle_z + \alpha_3 |3\rangle_z + \alpha_3^2 |5\rangle_z \Big].
\end{eqnarray}

\section{Two-component system}

Consider measurements of the generalized Pauli operators on a system composed of two subsystems, 
of dimension $d = d_1 d_0$.
The subsystems are described in Hilbert spaces of dimensions $d_1$ and $d_0$.
We first present a parameterization of states of subsystems into which the eigenvectors are decomposed,
and next utilize it to describe the structure of the eigenbases.
It is proven that all of the eigenvectors are either
entangled or disentangled.
Thus, one identifies the operators which can be measured on individual
subsystems (with additional feed-forward),
and those which require joint measurements.

Recall that, according to Eq. (\ref{STATES_INVOLVED}),
each eigenvector $| j_{g,a} \rangle$ involves every $f$th state of the $S_z$ basis,
$|a + \eta' f \rangle_z$, with $\eta' = 0,...,d/f-1$.
In turn, each of these states can be written in terms of subsystems, 
using base-$d_1d_0$ representation:
\begin{equation}
|a + \eta' f \rangle_z  = |d_0 \kappa_1 + \kappa_0 \rangle_z = |\kappa_1 \rangle_1 |\kappa_0 \rangle_0,
\end{equation}
where $\kappa_0 = [a + \eta' f]_{d_0}$ 
(the symbol $[x]_{d_0}$ denotes $x$ modulo $d_0$), 
and $\kappa_1= \lfloor (a + \eta' f)/d_0 \rfloor = 0,...,d_1-1$,
where $\lfloor x \rfloor$ denotes an integer part of $x$.
The number of distinct states of subsystem ``0''
is given by the number of different values of $\kappa_0$.
Since other values of $a$ only shift $\kappa_0$,
leaving the number of distinct values unaffected, one can put $a=0$, and thus $\kappa_0 = [\eta' f]_{d_0}$.
To calculate $\kappa_0$'s one divides $f$ by $d_0$,
and denotes the integer part of this division by $\xi$.
Thus, one can write $f/d_0 = \xi + p/d_0$.
The integers $p$ and $d_0$ can have common factors,
and the fraction $p/d_0$ may be simplified to
an irreducible form $P/D_0$.
Thus,
for $\eta'=0$ and $\eta'=D_0$ (and any multiple of $D_0$)
 $\kappa_0$ equals zero.

For $\eta' > D_0$ the values of $\kappa_0$ repeat themselves.
If one takes an integer $x$ and computes the value of $\kappa_0$ for $\eta'=D_0+x$:
$\kappa_0 = [(D_0+x)f]_{d_0} = [[D_0f]_{d_0}+[xf]_{d_0}]_{d_0} = [xf]_{d_0}$,
it is the same as for $\eta'=x$
(we have used the properties of addition in modulo calculus).
Thus there are $D_0$ different values of $\kappa_0$, 
or orthogonal states $|\kappa_0 \rangle_0$ in the decomposition of every $| j_{g,a} \rangle$.

Moreover, for $\eta' = d/f = d_1d_0/f$ the value of $\kappa_0$ again equals zero,
i.e. each state $|\kappa_0 \rangle_0$ (one of $D_0$ distinct states) 
appears in $| j_{g,a} \rangle$
exactly the same number of times.
This gives the number of orthogonal states $|\kappa_1 \rangle_1$ 
associated with any given $|\kappa_0 \rangle_0$, which will be denoted as $D_1$.
Since for different $\eta'$ vectors $|a + \eta' f \rangle_z$ are orthogonal, 
the states of subsystem ``1'', associated with the same $|\kappa_0 \rangle_0$
must be orthogonal.
Notice that $d/f$ factorizes into 
$d/f = D_1D_0$,
and this is a general property of an $S_{kl}$ operator.

One can introduce an integer $K_0 = 0,...,D_0-1$ 
to enumerate distinct states of subsystem ``0''.
In a similar way, for a fixed state $|\kappa_0 \rangle_0$,
one can enumerate orthogonal states of subsystem ``1''
with an integer $K_1=0,...,D_1-1$.
Since $\eta=0,...,D_1D_0-1$, it can be decomposed within the new variables $K_0$ and $K_1$ as:
\begin{equation}
\eta = D_0 K_1 + K_0.
\label{ETA}
\end{equation}
Within this decomposition every state $| a + \eta k \rangle_z$, 
into which the eigenvectors are decomposed, (\ref{EIGENVECTORS}),
can be written as $|a + K_0 k + D_0 k K_1 \rangle_z$.
To find its base-$d_1d_0$ form one needs to divide $a + K_0 k + D_0 k K_1$ by $d_0$,
and extract integer and modulo parts.
Since $k=wf$ one finds that 
$D_0 k/d_0 = w D_0 f/d_0 = w D_0 \xi + P$
is an integer, or equivalently $D_0 k$ is a multiple of $d_0$.
That is, $D_0 k$ does not contribute to the modulo part,
and one can write:
\begin{equation}
| a + \eta k \rangle_z = | \lfloor (a + k K_0)/d_0 \rfloor + \frac{D_0}{d_0} k K_1 \rangle_1 | a + k K_0\rangle_0.
\label{EIGEN_Z_SUBST}
\end{equation}

Let us summarize the parameterization just described. 
In the decomposition of any $| j_{g,a} \rangle$
one finds $D_0$ distinct states of 
subsystem ``0'', $|\kappa_0(a,K_0) \rangle_0 \equiv | a + k K_0\rangle_0$, 
with $K_0 = 0,...,D_0-1$.
In turn, for a fixed $K_0$, 
there are $D_1$ distinct states of subsystem ``1'', 
$|\kappa_1(a,K_1,K_0) \rangle_1 \equiv | \lfloor (a + k K_0)/d_0 \rfloor + \frac{D_0}{d_0} k K_1 \rangle_1$.

Within this parameterization any eigenstate has the following form:
\begin{equation}
| j_{g,a} \rangle = \frac{1}{\sqrt{D_1D_0}} 
\sum_{K_0=0}^{D_0-1} \sum_{K_1=0}^{D_1-1} c_{(D_0 K_1 + K_0)k,j_{g,a}}
| \kappa_1(a,K_1,K_0) \rangle_1 | \kappa_0(a,K_0) \rangle_0,
 \label{EIGEN_DECOMP_SUBSYST}
\end{equation}
where the coefficients $c_{(D_0 K_1 + K_0)k,j_{g,a}}$ denote the phase of 
coefficients in Eq. (\ref{EIGENVECTORS}).

To understand the structure of the eigenbasis 
take for a fixed state
$|\kappa_0(a,K_0) \rangle_0$ in Eq. (\ref{EIGEN_DECOMP_SUBSYST}),
the state of subsystem ``1'' with which it is associated, namely:
\begin{equation}
| \psi_{\kappa_0,j_{g,a}} \rangle_1 = 
\frac{1}{\sqrt{D_1}} \sum_{K_1 = 0}^{D_1-1}
c_{(D_0 K_1 + K_0)k,j_{g,a}} | \kappa_1(a,K_1,K_0) \rangle_1.
\end{equation}
It will be shown that within the same eigenvector $| j_{g,a} \rangle$,
any two states  $| \psi_{\kappa_0,j_{g,a}} \rangle_1$
and $| \psi_{\kappa_0',j_{g,a}} \rangle_1$,
for $\kappa_0 \ne \kappa_0'$,
are either \emph{orthogonal} or \emph{the same}.
A similar result holds for the states of different eigenvectors,
with the same value of $\kappa_0$.

Let us first consider states of subsystem ``1''
within the same eigenvector.
Their scalar product, $_1 \langle \psi_{\kappa_0',j_{g,a}}| \psi_{\kappa_0,j_{g,a}} \rangle_1$,
is given by:
\begin{eqnarray}
&&\sum_{K_1' = 0}^{D_1-1} \sum_{K_1 = 0}^{D_1-1}
c_{(D_0 K_1' + K_0')k,j_{g,a}}^{*} 
c_{(D_0 K_1 + K_0)k,j_{g,a}} \textrm{ }_1\langle \kappa_1(a,K_1',K_0')| \kappa_1(a,K_1,K_0) \rangle_1.
\end{eqnarray}
Since for different $\kappa_0$'s the states $| \kappa_1(a,K_1,K_0) \rangle_1$
are shifted,
the scalar product $_1\langle \psi_{\kappa_0',j_{g,a}}| \psi_{\kappa_0,j_{g,a}} \rangle_1$
is either equal to zero,
if the individual states involved are orthogonal,
or it is equal to the sum of $D_1$ terms:
$\sum_{K_1=0}^{D_1-1}
c_{(D_0 K_1 + K_0')k,j_{g,a}}^{*} c_{(D_0 K_1 + K_0)k,j_{g,a}}$.
Using explicit form of the coefficients
one finds that 
the scalar product is proportional to:
\begin{equation}
_1\langle \psi_{\kappa_0',j_{g,a}} |\psi_{\kappa_0,j_{g,a}} \rangle_1
\sim
\frac{1}{D_1} \sum_{K_1=0}^{D_1-1} \alpha_{d}^{K_1 klD_0(K_0-K_0')},
\end{equation}
where the only relevant terms are given, 
involving in the exponent the products of $K_1$ and $K_0$.
Since  $k=wf$
and $d = D_1 D_0 f$,
right-hand side equals to the Kronecker delta:
\begin{equation}
_1\langle \psi_{\kappa_0',j_{g,a}} |\psi_{\kappa_0,j_{g,a}} \rangle_1
\sim
\delta_{wl(K_0-K_0'),\mu D_1},
\end{equation}
with $\mu = 0,1,2,...$.
The states of subsystem ``1''
are either orthogonal or the same 
(up to a global phase, which can be put to multiply them).
If the Kronecker delta is equal to one, one has
$|\psi_{\kappa_0,j_{g,a}} \rangle_1 = e^{i \phi} |\psi_{\kappa_0',j_{g,a}} \rangle_1$,
i.e. a vector of subsystem ``1''
is multiplied by a superposition of
corresponding $|\kappa_0\rangle_0$'s,
with coefficients defined by (\ref{EIGEN_DECOMP_SUBSYST}),
respectively multiplied by the phase $e^{i \phi}$.
Since different $|\kappa_0\rangle_0$'s are orthogonal, 
every vector $|j_{g,a} \rangle$ can be written
as a superposition of bi-orthogonal product states.
In other words, one has a Schmidt decomposition
of the eigenvectors.

Moreover, for different eigenvectors, 
the states $|\psi_{\kappa_0,j_{g,a}} \rangle_1$ and $|\psi_{\kappa_0,j_{g',a'}} \rangle$,
which correspond to the same state of subsystem ``0'',
are also either orthogonal or the same.
Their scalar product $_1 \langle \psi_{\kappa_0,j_{g',a'}} |\psi_{\kappa_0,j_{g,a}} \rangle_1$
involves scalar products $_1\langle \kappa_1(a',K_1',K_0)| \kappa_1(a,K_1,K_0) \rangle_1$.
For different 
eigenvectors the states $| \kappa_1(a,K_1,K_0) \rangle_1$
can be shifted,
and one has:
\begin{equation}
_1 \langle \psi_{\kappa_0,j_{g',a'}} |\psi_{\kappa_0,j_{g,a}} \rangle_1
\sim
\frac{1}{D_1} \sum_{K_1=0}^{D_1-1} [\lambda_{j_{g',0}}\lambda_{j_{g,0}}^{-1}]^{K_1 D_0}.
\end{equation}
Since the product of eigenvalues, $\lambda_{j_{g',0}}\lambda_{j_{g,0}}^{-1}$, 
is equal to $\alpha_{d/f}^{g'-g} = \alpha_{D_1D_0}^{g'-g}$, one has:
\begin{equation}
_1 \langle \psi_{\kappa_0,j_{g',a'}} |\psi_{\kappa_0,j_{g,a}} \rangle_1
\sim
\frac{1}{D_1} \sum_{K_1=0}^{D_1-1} \alpha_{D_1}^{K_1 (g'-g)} = \delta_{g'-g,\mu D_1},
\end{equation}
with $\mu = 0,1,2,...$.
Notice that the last Kronecker delta does not depend on $\kappa_0$.
E.g., if for some $\kappa_0$ one finds that vectors $|\psi_{\kappa_0,j_{g',a'}} \rangle_1$
and $|\psi_{\kappa_0,j_{g,a}} \rangle_1$ are orthogonal, then
the same relation holds for any other $\kappa_0$,
i.e. all the eigenstates have exactly the same number of terms in the Schmidt form
(the same Schmidt number).
If $\delta_{g'-g,\mu D_1} = 1$ all the states $|\psi_{\kappa_0,j_{g,a}} \rangle_1$
in the decomposition of $|j_{g,a} \rangle$ are the same (up to a global phase)
as those entering $|j_{g',a'} \rangle$.
In this case the coefficients which multiply products 
$|\psi_{\kappa_0,j_{g,a}} \rangle_1 |\kappa_0(a,K_0) \rangle_0$
make the two eigenvectors
orthogonal.

To conclude,
given that only individual measurements on subsystems 
are available to an experimenter,
she/he can learn from above considerations
whether it is possible
to measure a generalized Pauli operator defined
on the whole system (of two components).

\section{Two-Bases quantum cryptography}

Let us apply the developed formalism.
Consider a two-bases quantum cryptography protocol
with higher-dimensional systems, as described in Ref. \cite{MOHAMED}. One has a qudit  randomly prepared in a state of a certain basis, or of another basis,  which is unbiased with respect to the first one 
\cite{MUBS,MUBS2}.
The measurement basis is also randomly chosen between these two \footnote{The performance of the two-bases protocol
is only slightly worse than the performance of a many-bases protocol
(compare Table I of Ref. \cite{MOHAMED}).}.
Interestingly, if a qudit is composed of two subsystems, 
the measurements involved in the protocol do not require
any joint actions.

The two mutually unbiased bases 
can be chosen as the eigenbases of $S_z$ and $S_x$ operators.
Using the above construction to $S_x = S_{10}$ 
one immediately finds, for arbitrary dimension, 
the well-known Fourier relation between the $S_z$ and $S_x$ eigenbases:
\begin{equation}
|j\rangle_x = \frac{1}{\sqrt{d}} \sum_{\kappa = 0}^{d-1} \alpha_d^{- \kappa j} |\kappa \rangle_z.
\end{equation}
Let us define the eigenbasis of a global $S_z$ operator as:
\begin{equation}
| \kappa \rangle_z = | d_0 \kappa_1 + \kappa_0 \rangle_z \equiv |\kappa_1 \rangle_1 |\kappa_0 \rangle_0,
\label{DEFINITION}
\end{equation}
where $\kappa_i = 0,...,d_i - 1$, and $|\kappa_0 \rangle_0$, $|\kappa_1 \rangle_1$
denote the states of subsystems ``0'' and ``1'', respectively.
Within this definition a measurement of the global observable $S_z$
is equivalent to individual measurements on the components.
These individual measurements reveal the values of $\kappa_0$ and $\kappa_1$,
and the eigenvalue of $S_z$ is $\alpha_d^{d_0 \kappa_1 + \kappa_0}$
[due to Eq. (\ref{S_DEFINITION})].

To measure $S_x$ one uses the definition (\ref{DEFINITION}) and the fact that $d=d_1d_0$,
and finds that:
\begin{equation}
| j \rangle_x = \frac{1}{\sqrt{d_1}} 
\sum_{\kappa_1=0}^{d_1-1} \alpha_d^{-d_0 \kappa_1 j} |\kappa_1 \rangle_1 \otimes
\frac{1}{\sqrt{d_0}} 
\sum_{\kappa_0=0}^{d_0-1} \alpha_d^{-\kappa_0 j}|\kappa_0 \rangle_0,
\label{SEPARATED}
\end{equation}
where we have used the symbol $\otimes$
to stress the factorization of this state.
For $j = j_1 + d_1 j_0$
the state of subsystem ``1'' reads:
\begin{equation}
\frac{1}{\sqrt{d_1}}  \sum_{\kappa_1=0}^{d_1-1} \alpha_d^{-d_0 \kappa_1 j_1-d_0 \kappa_1 d_1 j_0} |\kappa_1 \rangle_1.
\end{equation}
Since $\alpha_d^{d_0} = \alpha_{d_{1}}$, see (\ref{ALPHA}),
and $e^{-i 2\pi \kappa_1 j_0} = 1$ a
measurement on this subsystem in the basis:
\begin{equation}
|\phi_{j_1} \rangle_1 = \frac{1}{\sqrt{d_1}}  \sum_{\kappa_1=0}^{d_1-1} \alpha_{d_1}^{-\kappa_1 j_1} |\kappa_1 \rangle_1,
\end{equation}
reveals the value of $j_1$.
The value of $j_0$ can be measured once $j_1$ is known.
A measurement in the basis:
\begin{equation}
|\psi_{j_0} \rangle_0 = \frac{1}{\sqrt{d_0}}  \sum_{\kappa_0=0}^{d_0-1} \alpha_{d}^{-(j_1 + d_1 j_0) \kappa_0} |\kappa_0 \rangle_0,
\end{equation}
on the subsystem ``0''
reveals the value of $j_0$.
In this way all values of $j$ can be measured
using individual measurements only,
where the measurement on subsystem ``0'' depends on 
the outcome of the measurement on subsystem ``1''
(feed-forward technique).

\section{Quantum tomography}

Another application utilizes the fact that the $S_{kl}$ operators form
a basis in a Hilbert-Schmidt space, and thus
can be used in quantum tomography.
Quantum tomography (reconstruction of a density matrix)
aims at an estimation of an unknown quantum state.
The tomography of qubits was described in \cite{JAMES}.
Soon after, the generalization
to higher-dimensional systems was given in \cite{THEW}.
The approach described there is based on Hermitian operators.
Here we follow the unitary operators approach,
and explicitly present, in the next section, 
suitable devices to perform tomography
of polarisation-path qudits.

Since $d^2$ qudit operators $S_{kl}$ form a basis in the 
Hilbert-Schmidt space,
they uniquely describe an arbitrary state of a qudit:
\begin{equation}
\rho = \frac{1}{d} \sum_{k,l=0}^{d-1} s_{kl} S_{kl},
\end{equation}
where $s_{00}=1$ for normalisation
as all $S_{kl}$ operators are traceless, except the identity. 
Tomography means to establish (measure) all of the $s_{kl}$ coefficients.
Since the $S_{kl}$ operators have the spectral decomposition
$S_{kl} = \sum_{j=0}^{d-1} \lambda_{j} |j\rangle \langle j |$,
the coefficients $s_{kl}$ can be written as:
\begin{equation}
s_{kl} = \textrm{Tr}(S_{kl}^{\dagger} \rho)
= \sum_{j=0}^{d-1} \lambda_{j}^* \textrm{Tr}(|j\rangle \langle j | \rho).
\label{TOMOGRAPHY}
\end{equation}
The eigenvectors $| j \rangle$ form an orthonormal set,
and the trace gives the probability, $p_{j}$, 
to obtain the $j$th outcome
in the measurement of $S_{kl}$ on the system prepared in the state $\rho$.
Finally, to perform tomography one needs to build the devices
capable to measure $S_{kl}$, 
and collect data to estimate probabilities (relative frequencies) of different outcomes, $p_{j}$.
We focus on measurement devices for polarisation-path qudits.

\section{Polarisation-path qudits}

Although the general requirements for a measurement
involve feed-forward and joint operations on subsystems,
there are certain physical realisations of composite qudits 
which incorporate these requirements in a simple way.
The polarisation-path qudit is an example.
There, a qudit is encoded in a polarized photon,
which has many possible propagation paths \footnote{Note that only qudits of an even dimension can be realized in this way.}.
First, we explicitly present devices capable to measure
all $S_{kl}$ operators in the simplest case of two paths.
Next, the setups for any number of paths are discussed.

Consider a polarized photon with two accessible paths.
Its state is described in a four dimensional Hilbert space,
i.e. there are $15$ different $S_{kl}$ operators to measure
(we put $s_{00} = 1$ from the very definition).
However, some of them commute (contrary to the qubit case) 
and the measurement of one of them reveals the values of the others.

From the definition, the eigenstates of $S_z$ are given by:
\begin{eqnarray}
|0\rangle_z = |0_z\rangle_1 |0_z\rangle_0, & \quad & |1\rangle_z = |0_z\rangle_1 |1_z\rangle_0, \nonumber \\
|2\rangle_z = |1_z\rangle_1 |0_z\rangle_0, & \quad & |3\rangle_z = |1_z\rangle_1 |1_z\rangle_0,
\label{QUQUAT}
\end{eqnarray}
where subsystem ``0'' is a polarisation of a photon,
and subsystem ``1'' is a path. 
E.g. $|2 \rangle_z = |1_z\rangle_1 |0_z\rangle_0$
denotes a horizontally polarised photon in the path $|1_z\rangle_1$.
The $z$ index inside the two-level kets 
denotes the fact that they are chosen as
the eigenstates of the individual $\sigma_z^{(n)}$ operators,
i.e. $\sigma_z^{(n)} |b_z \rangle_n = (-1)^b |b_z \rangle_n$.
The device that measures $S_z$ simply checks
which polarisation a photon has in a certain path.
This can easily be achieved with polarizing beam-splitters.
Moreover, the same device also measures the values of $S_z^2$ and $S_z^3$,
as these operators commute with $S_z$. Their eigenvalues
are powers of the $S_z$ eigenvalues.
Interestingly, the observables $S_{21}$ and $S_{23}$ can be measured in a similar way.
After expressing the eigenvectors of, say, $S_{21}$ in the $|\kappa \rangle_z$ basis,
and with definitions (\ref{QUQUAT}), one finds:
\begin{eqnarray}
|0 \rangle = |0_y\rangle_1 |1_z\rangle_0, & \quad & |1 \rangle = |0_y\rangle_1 |0_z\rangle_0, \nonumber \\
|2 \rangle = |1_y\rangle_1 |1_z\rangle_0, & \quad & |3 \rangle = |1_y\rangle_1 |0_z\rangle_0,
\end{eqnarray}
where $|b_y\rangle_n$ is the eigenbasis of the individual $\sigma_y^{(n)}$ operator,
$|b_y\rangle_n = \frac{1}{\sqrt{2}}(|0_z\rangle_n + i(-1)^b |1_z \rangle_n)$.
To measure this observable the paths meet on a beam-splitter
(which gives a phase $\frac{\pi}{2}$ to the reflected beam)
where different eigenstates $|b_y\rangle_1$ are directed
into different output ports,
followed by polarizing beam-splitters.

\begin{figure}[t]
\begin{center}
\includegraphics{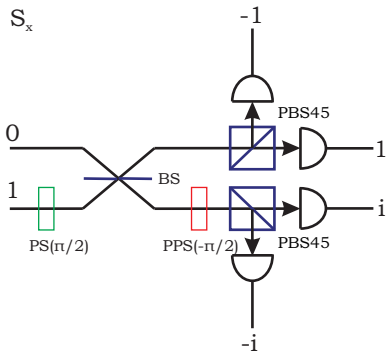}
\end{center}
\caption{Feed-forward is not needed for polarization-path
qudits -- it is automatically realized behind the beam-splitter.
This setup measures the operator $S_x$, for $d=4$.
The $\frac{\pi}{2}$ phase shift (PS($\pi/2$)) in the path $|1_z\rangle_1$ and the beam-splitter (BS)
perform the path measurement, $\sigma_x^{(1)}$.
The path state $| 0_x \rangle_1$ goes to the upper arm where the polarisation
is measured in the $\sigma_x^{(0)}$ basis with the polarizing beam-splitter
which transmits $| 0_x \rangle_0$ (denoted as PBS45).
In case of the path state $| 1_x \rangle_1$ the photon goes to the lower arm,
where its $| 1_z \rangle_0$ polarisation component is phase shifted by $-\frac{\pi}{2}$ 
(PPS($-\pi/2$)).
Next, the photon enters PBS45, and is detected in one of its outputs.
The eigenvalues corresponding to clicks of each detector are also written.}
\label{SX}
\end{figure}

The $S_x = S_{10}$ observable (and its powers) can be measured
individually with an additional feed-forward.
Also $S_{12}$ and $S_{32}$ are measurable in this way.
To see how the feed-forward method is realized, 
let us study the $S_x$ observable.
Its eigenvectors read:
\begin{eqnarray}
|0 \rangle = |0_x\rangle_1 |0_x\rangle_0, & \quad & |1 \rangle = |1_x\rangle_1 |0_y\rangle_0, \nonumber \\
|2 \rangle = |0_x\rangle_1 |1_x\rangle_0, & \quad & |3 \rangle = |1_x\rangle_1 |1_y\rangle_0,
\label{SX_EIGEN}
\end{eqnarray}
where the index $x$ denotes the eigenbasis of the individual $\sigma_x^{(n)}$ operator, given by
$| b_x \rangle_n = \frac{1}{\sqrt{2}}(| 0_z \rangle_n + (-1)^b | 1_z \rangle_n)$.
Depending on the outcome of the path measurement in the $\sigma_x^{(1)}$ basis,
polarisation is measured in the $\sigma_x^{(0)}$ or $\sigma_y^{(0)}$ basis.
However (here comes the beauty of the approach utilizing the paths),
appropriate phase and a beam-splitter
drive different $\sigma_x^{(1)}$ path eigenstates
into different output ports of the beam-splitter.
In this way feed-forward is not needed.
It is now enough to put polarisation checking devices behind the proper
outputs of the beam-splitter (see Fig. \ref{SX}).

\begin{figure}[t]
\begin{center}
\includegraphics{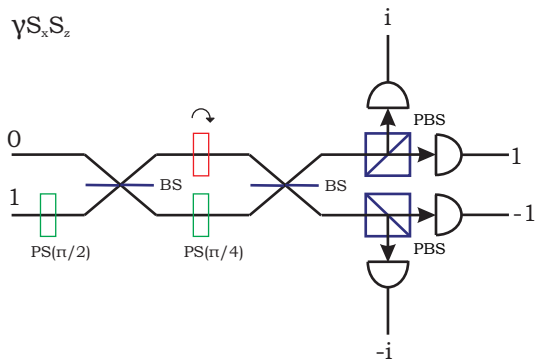}
\end{center}
\caption{Mach-Zehnder interferometer, with a polarisation rotator in one arm,
followed by polarizing beam-splitters, is the most advanced device used in measurements
of $S_{kl}$, for $d=4$.
This setup, which measures the operator $\gamma S_x S_z$ (with $\gamma = e^{i \pi/4}$),
distinguishes maximally entangled states of paths and polarisations.
First, with the $\frac{\pi}{2}$ phase shift (PS($\pi/2$)) and the beam-splitter (BS), 
the $\sigma_x^{(1)}$ eigenstates
are converted into $\sigma_z^{(1)}$ eigenstates. 
Next, the $\pi/4$ phase (PS($\pi/4$))
is applied in the lower arm, where $| 1_x \rangle_1$ is directed.
In the upper arm polarization is rotated (with the plate $\curvearrowright$),
such that in both arms it is the same.
Finally, specific clicks behind the beam-splitter and polarising beam-splitters
distinguish the states (\ref{SXSZ_EIGEN}).}
\label{SXSZ}
\end{figure}

The eigenstates of 
the last five observables are maximally entangled states of subsystems.
Some of these observables, to keep the spectrum in the domain of fourth roots of unity,
need to be multiplied by $\gamma \equiv \alpha_4^{1/2} = e^{i \pi/4}$.
Take as an example $S_{11}$ operator in the form $S_{11} = \gamma S_x S_z$. 
Its eigenstates are given by:
\begin{eqnarray}
|0 \rangle &=& (1/\sqrt{2})\Big[|0_x\rangle_1 |1_z\rangle_0 - i\gamma |1_x\rangle_1 |0_z\rangle_0 \Big], \nonumber \\
|1 \rangle &=& (1/\sqrt{2})\Big[|0_x\rangle_1 |0_z\rangle_0 - i\gamma |1_x\rangle_1 |1_z\rangle_0 \Big], \nonumber \\
|2 \rangle &=& (1/\sqrt{2})\Big[|0_x\rangle_1 |1_z\rangle_0 + i\gamma |1_x\rangle_1 |0_z\rangle_0 \Big], \nonumber \\
|3 \rangle &=& (1/\sqrt{2})\Big[|0_x\rangle_1 |0_z\rangle_0 + i\gamma |1_x\rangle_1 |1_z\rangle_0 \Big].
\label{SXSZ_EIGEN}
\end{eqnarray}
To distinguish between these states one needs to build an interferometer 
like the one in the Fig. \ref{SXSZ}.
The same setup measures $S_{22}$ and $S_{33}$,
which commute with $S_{11}$.
Finally, when different phase shifts are used, 
this setup also measures the remaining $S_{13}$ and $S_{31}$ observables.

To sum up, the most involved device,
used in the measurements of generalized Pauli operators on a composite qudit encoded in two paths
and polarization of a photon, involves Mach-Zehnder interferometer
with a polarization rotator in one arm, followed by polarizing beam-splitters (Fig. \ref{SXSZ}).
Most of the observables are realizable with a single beam-splitter
followed by polarizing beam-splitters.

Generally, it is possible to perform arbitrary $S_{kl}$ measurement on
polarized photons with many, $d_1$, accessible paths.
With polarising beam-splitters in each propagation path
one transforms initial polarisation-path state $| j \rangle$
into a double-number-of-paths state $|p \rangle$, in $2d_1$ dimensional Hilbert space
(each polarising beam-splitter generates two distinct spatial modes).
According to Ref. \cite{RECK}
one can always realize a unitary which
brings the states $|p \rangle$
to the states of well-defined propagation direction.
Thus, $2d_1$ detectors monitoring these final paths distinguish all
the eigenvectors $| j \rangle$.

\section{Conclusions}

Higher-dimensional quantum systems can find many applications,
both in foundations of physics and in applied quantum information.
A method of construction of qudits, studied here, 
is to compose them
of other, lower dimensional, subsystems.
In such a case, if a global observable has some entangled eigenvectors,
its measurement naturally requires joint actions on subsystems.
If eigenvectors factorize,
the observable is measurable individually,
sometimes with an additional feed-forward.
Thus, in order to design a setup
capable to measure an observable, 
its eigenproblem must be solved.

Here, the eigenproblem of the unitary generalizations
of Pauli operators is solved, for arbitrary dimensions,
and Schmidt decomposition
of the eigenvectors,
for qudits composed of two components, is derived.
Using these results 
quantum cryptography with two bases, 
operating on a two-component qudit,
is shown not to involve any joint measurements.

Finally, simple optical devices, 
capable to measure all
generalized Pauli operators on polarisation-path qudits,
are presented.
These experimentally feasible devices allow full state tomography.
In case of two different paths,
the most complicated device is a Mach-Zehnder interferometer,
with a polarisation rotator in one arm, 
followed by polarizing beam-splitters.

\section{Acknowledgements}

The author is extremely grateful to Professor Marek \.Zukowski
for useful comments.
Marcin Wie\'sniak is also gratefully acknowledged.
The work is part of the MNiI Grant No. 1 P03B 049 27 and the $6$th EU Framework programme QAP (Qubit Applications) Contract No. 015848.
The author is supported by the Foundation for Polish Science.


\begin{thebibliography}{00}

\bibitem{1PERES}
H. Bechmann-Pasquinucci and A. Peres,
Phys. Rev. Lett. {\bf 85}, 3313 (2000).
\bibitem{1BRUS}
D. Bru{\ss} and C. Macchiavello,
Phys. Rev. Lett. {\bf 88}, 127901 (2002).

\bibitem{MOHAMED}
N. J. Cerf, M. Bourennane, A. Karlsson, and N. Gisin,
Phys. Rev. Lett. {\bf 88}, 127902 (2002).

\bibitem{PRL_CCP}
{\v C}. Brukner, M. \.Zukowski, and A. Zeilinger, 
Phys. Rev. Lett. \textbf{89}, 197901 (2002).
\bibitem{IJQI_CCP}
{\v C}. Brukner, T. Paterek, M. \.Zukowski,
Int. J. Quant. Inf. {\bf 1}, 519 (2003).

\bibitem{1MULTIPATH}
A. Zeilinger, H. J. Bernstein, D. M. Greenberger, M. A. Horne, and M. \.Zukowski
in \textit{Quantum Control and Measurement}, 
edited by H. Ezawa and Y. Murayama (Elsevier, Amsterdam, 1993).
\bibitem{2MULTIPATH}
A. Zeilinger, M. \.Zukowski, M. A. Horne, H. J. Bernstein, and D. M. Greenberger
in {\it Quantum Interferometry},
edited by F. DeMartini and A. Zeilinger (World Scientific, Singapore, 1994).

\bibitem{HYPERENTANGLEMENT}
J. T. Barreiro, N. K. Langford, N. A. Peters, and P. G. Kwiat,
Phys. Rev. Lett. {\bf 95}, 260501 (2005).

\bibitem{PROPOSAL}
Z.-B. Chen, J.-W. Pan, Y.-D. Zhang, {\v C}. Brukner, and A. Zeilinger,
Phys. Rev. Lett. {\bf 90}, 160408 (2003).
\bibitem{1EXP}
T. Yang, Q. Zhang, J. Zhang, J. Yin, Z. Zao, M. \.Zukowski, Z.-B. Chen, and J.-W. Pan,
Phys. Rev. Lett. {\bf 95}, 240406 (2005).
\bibitem{2EXP}
C. Cinelli, M. Barbieri, R. Perris, P. Mataloni, and F. De Martini,
Phys. Rev. Lett. {\bf 95}, 240405 (2005).

\bibitem{SCHWINGER}
J. Schwinger,
Proc. Nat. Ac. Sci. {\bf 46}, 570 (1960).
\bibitem{FIVEL}
D. I. Fivel, 
Phys. Rev. Lett. \textbf{74}, 835 (1995).
\bibitem{GOTTESMAN}
D. Gottesman,
in \emph{Quantum Computing and Quantum Communications:
First NASA International Conference}, 
edited by C. P. Williams (Springer-Verlag, Berlin, 1999).
\bibitem{PR}
A. O. Pittenger and M. H. Rubin, 
Phys. Rev. A \textbf{62}, 32313 (2000).

\bibitem{1BELL}
N. J. Cerf, S. Massar, and S. Pironio,
Phys. Rev. Lett. {\bf 89}, 80402 (2002).
\bibitem{2BELL}
W. Son, J. Lee, and M. S. Kim,
Phys. Rev. Lett. {\bf 96}, 60406 (2006).
\bibitem{3BELL}
J. Lee, S.-W. Lee, and M. S. Kim,
Phys. Rev. A {\bf 73}, 32316 (2006).

\bibitem{NIELSEN}
M. A. Nielsen, M. J. Bremner, J. L. Dodd, A. M. Childs, and C. M. Dawson,
Phys. Rev. A {\bf 66}, 22317 (2002).

\bibitem{SPECTRAL_QUDITS}
A. Muthukrishnan and C. R. Stroud, Jr.,
Phys. Rev. A {\bf 62}, 52309 (2000).

\bibitem{MUBS}
W. K. Wooters and B. D. Fields, 
Ann. Phys. (N. Y.) {\bf 191}, 363 (1989).

\bibitem{MUBS2}
S. Bandyopadhyah, P. O. Boykin, V. Roychowdhury, and F. Vatan,
Algorithmica {\bf 34}, 512 (2002).

\bibitem{JAMES}
D. F. V. James, P. G. Kwiat, W. J. Munro, and A. G. White,
Phys. Rev. A {\bf 64}, 52312 (2001).

\bibitem{THEW}
R. T. Thew, K. Nemoto, A. G. White, and W. J. Munro,
Phys. Rev. A {\bf 66}, 12303 (2002).

\bibitem{RECK}
M. Reck, A. Zeilinger, H. J. Bernstein, and P. Bertani,
Phys. Rev. Lett. {\bf 73}, 58 (1994).


\end{thebibliography}
\end{document}